\documentclass[twocolumn]{aastex6}
\usepackage{graphicx}
\usepackage{amssymb,amsfonts,amsmath,amstext,amsgen,amsopn,amsxtra,indentfirst,times}

\begin{document}

\title{A Cloudiness Index for Transiting Exoplanets Based on the Sodium and Potassium Lines: Tentative Evidence for Hotter Atmospheres Being Less Cloudy at Visible Wavelengths}

\author{Kevin Heng\altaffilmark{1}}
\altaffiltext{1}{University of Bern, Center for Space and Habitability, Sidlerstrasse 5, CH-3012, Bern, Switzerland.  Email: kevin.heng@csh.unibe.ch}

\begin{abstract}
We present a dimensionless index that quantifies the degree of cloudiness of the atmosphere of a transiting exoplanet.  Our cloudiness index is based on measuring the transit radii associated with the line center and wing of the sodium or potassium line.  In deriving this index, we revisited the algebraic formulae for inferring the isothermal pressure scale height from transit measurements.  We demonstrate that the formulae of Lecavelier et al. and Benneke \& Seager are identical: the former is inferring the temperature while assuming a value for the mean molecular mass and the latter is inferring the mean molecular mass while assuming a value for the temperature.  More importantly, these formulae cannot be used to distinguish between cloudy and cloudfree atmospheres.  We derive values of our cloudiness index for a small sample of 7 hot Saturns/Jupiters taken from Sing et al.  We show that WASP-17b, WASP-31b and HAT-P-1b are nearly cloudfree at visible wavelengths.  We find the tentative trend that more irradiated atmospheres tend to have less clouds consisting of sub-micron-sized particles.  We also derive absolute sodium and/or potassium abundances $\sim 10^2$ cm$^{-3}$ for WASP-17b, WASP-31b and HAT-P-1b (and upper limits for the other objects).  Higher-resolution measurements of both the sodium and potassium lines, for a larger sample of exoplanetary atmospheres, are needed to confirm or refute this trend.
\end{abstract}


\section{Introduction}

As the atmospheres of more exoplanets are being characterized, astronomers are finding that a non-negligible fraction of them are cloudy (e.g., \citealt{deming13,knutson14,kreidberg14,sing16}).  Cloudy atmospheres are challenging to decipher, because the interpretation of their chemical abundances is degenerate with the degree of cloudiness (e.g., \citealt{lee13}).  It motivates the construction of dimensionless indices for quantifying the degree of cloudiness of the atmosphere of a transiting exoplanet.  If trends between these indices and the properties of the exoplanet or its star are found, then one may select a sub-sample of cloudfree objects for more indepth atmospheric characterization using future facilities such as the James Webb Space Telescope.  Essentially, any trends found allow us to perform triage on the cloudy objects.

\cite{stevenson16} has previously defined a cloudiness index based on near-infrared diagnostics: the strength of the water feature at about 1.4 $\mu$m, as probed by WFC3 on the Hubble Space Telescope, and the J band.  The study of \cite{stevenson16} found that exoplanets with equilibrium temperatures higher than 700 K and surface gravities greater than $\log{g}=2.8$ (cgs units) are more likely to be cloudfree.  \cite{sing16} found a correlation between the strength of the 1.4 $\mu$m water feature and the difference in transit radii between the near- and mid-infrared wavelengths.  The approach of \cite{sing16} is based on using one-dimensional, self-consistent model atmospheres in chemical and radiative equilibrium, along with an assumed value for the metallicity, as a baseline for defining what ``cloudfree" means.  

In the current Letter, we take a different and complementary approach by constructing a cloudiness index based on the notion that, in a cloudfree atmosphere, the difference in transit radii between the line center and wing of the sodium or potassium lines should be straightforwardly calculable.  A cloudy atmosphere would have a value of this difference in transit radii that is \textit{less} than the cloudfree value \citep{ss00}, thereby naturally allowing us to define a cloudiness index.  We apply our cloudiness index to the sample of exoplanets presented in \cite{sing16} and report a tentative trend between cloudiness at visible wavelengths and the strength of stellar irradiation.

\section{Revisiting Benneke \& Seager (2012)}

Before constructing our cloudiness index, we examine in detail if measurements of the spectral slope in the visible range of wavelengths may be used, in isolation, to quantify the degree of cloudiness of an atmosphere.  \cite{bs12} have previously published, in their appendix, approximate algebraic solutions to infer the mean molecular mass of an atmosphere based on a pair of transit measurements at different wavelengths.  We wish to point out that these formulae have an earlier origin and also elucidate the assumptions behind them.

In work preceding \cite{bs12}, \cite{lec08} have shown that, in an isothermal atmosphere, the temperature may be inferred from
\begin{equation}
T = \frac{mg}{k_{\rm B}} \frac{\partial R}{\partial \left( \ln{\sigma} \right)},
\label{eq:isotemp}
\end{equation}
where $m$ is the mean molecular mass, $g$ is the surface gravity, $k_{\rm B}$ is the Boltzmann constant, $R$ is the transit radius and $\sigma$ is the cross section for absorption or scattering.  Equation (\ref{eq:isotemp}) may be trivially re-arranged to yield equation (A5) of \cite{bs12},
\begin{equation}
m = \frac{k_{\rm B} T}{g} \left[ \frac{\partial R}{\partial \left( \ln{\sigma} \right)} \right]^{-1}.
\end{equation}

We now assume that the cross section is associated with scattering by \textit{both} molecules and aerosols/condensates, and that it has a power-law functional form,
\begin{equation}
\sigma = \left( A_{\rm molecules} + A_{\rm aerosols} \right) \lambda^\alpha,
\end{equation}
where $\lambda$ denotes the wavelength and $\alpha$ is a dimensionless index.  It immediately follows that
\begin{equation}
d\left( \ln{\sigma} \right) = \alpha ~d\left( \ln{\lambda} \right),
\end{equation}
independent of $A_{\rm molecules}$ and $A_{\rm aerosols}$.  Physically, it implies that measuring the spectral slope alone does not allow us to distinguish between scattering by molecules or aerosols, since the preceding expression has no dependence on either $A_{\rm molecules}$ or $A_{\rm aerosols}$.  Equation (\ref{eq:isotemp}) becomes \citep{lec08}
\begin{equation}
H = \frac{1}{\alpha} \frac{\partial R}{\partial \left( \ln{\lambda} \right)} \approx \frac{R_2 - R_1}{4 \ln{\left( \lambda_1/\lambda_2 \right)}},
\label{eq:scaleheight}
\end{equation}
where $R_1$ and $R_2$ correspond to the transit radii at the wavelengths $\lambda_1$ and $\lambda_2$, respectively.  The approximate expression in the preceding equation, which is equation (A6) of \cite{bs12}, derives from assuming that $\alpha=-4$ for Rayleigh scattering.  It is important to note that only $\partial R / \partial (\ln{\lambda})$ is an observable, which allows $\alpha H$ to be inferred.  It does not allow one to ascertain if Rayleigh scattering is at work, i.e., that $\alpha=-4$.  This is an assumption.  If we are associating the Rayleigh slope with aerosols/condensates, then we are assuming that the particles have a radius of $r \ll \lambda/2\pi$.

By further manipulating equation (\ref{eq:scaleheight}), we obtain equation (A7) of \cite{bs12},
\begin{equation}
m \approx \frac{4k_{\rm B} T}{g R_\star} \frac{\ln{\left( \lambda_1/\lambda_2 \right)}}{R_2/R_\star - R_1/R_\star}.
\label{eq:mmm}
\end{equation}
The preceding expression carries an additional assumption: the stellar radius ($R_\star$) is assumed to be the same at both wavelengths.

In summary, \cite{lec08} and \cite{bs12} are using the same algebraic formula to essentially infer the isothermal pressure scale height by assuming that Rayleigh scattering is at work ($\alpha=-4$).  \cite{lec08} then infer the temperature by assuming a value for the mean molecular mass\footnote{If $m_{\rm H}$ is the mass of the hydrogen atom, then the mean molecular weight is given by $\mu = m/m_{\rm H}$.}.  \cite{bs12} infer the mean molecular mass by assuming a value for the temperature.  From measuring the spectral slope alone, one cannot distinguish between Rayleigh scattering by molecules (e.g., hydrogen, nitrogen) or aerosols/condensates.

\begin{table*}
\label{tab:data}
\begin{center}
\caption{Observed Quantities for a Sample of Exoplanets}
\begin{tabular}{lcccccccc}
\hline
\hline
Name & $\lambda_0$ & $R_0/R_\star$ & $\lambda$ & $R/R_\star$ & $R_\star$ & $H_{\rm eq}$ & $g$ & $T_{\rm eq}$ \\
\hline
Physical units & ($\mu$m) & -- & ($\mu$m) & -- & ($R_\odot$) & (km) & (cm s$^{-2}$) & (K) \\
\hline
WASP-6b & 0.5893, 0.7684 & 0.14656(132), 0.14718(079) & 0.6059, 0.7299 & 0.14486(053), 0.14504(043) & 0.870 & 455 & 870 & 1150 \\
WASP-17b & 0.5893 & 0.13414(555) & 0.6124 & 0.11716(241) & 1.583 & 1662  & 360 & 1740 \\
WASP-31b & 0.7683 & 0.13338(200) & 0.7548 & 0.12452(112) & 1.12 & 1181 & 460 & 1580 \\	
WASP-39b & 0.5893, 0.7684 & 0.14977(229), 0.14828(290) & 0.5989, 0.7380 & 0.14603(118), 0.14438(066) & 0.918 & 940 & 410 & 1120 \\
HAT-P-1b & 0.5893, 0.76649 & 0.12109(146), 0.12680(120) & 0.6054, 0.7582 & 0.11778(050), 0.12480(140) & 1.115 & 605 & 750 & 1320 \\
HAT-P-12b & 0.5893, 0.7684 & 0.14182(203), 0.14443(269) & 0.6059, 0.7367 & 0.13937(105), 0.14085(131) & 0.701 & 590 & 560 & 960 \\
HD 189733b & 0.5895 & 0.15703(011) & 0.5980 & 0.15631(022) & 0.805 & 193 & 2140 & 1200 \\
\hline
\hline
\end{tabular}\\
Note: we use $R_\odot = 6.9566 \times 10^{10}$ cm.  We use the shorthand notation, e.g., $0.14656(132)$ means $0.14656 \pm 0.00132$.
\end{center}
\end{table*}

\begin{table*}
\label{tab:data2}
\begin{center}
\caption{Cloudiness Index and Absolute Sodium and Potassium Abundances for a Sample of Exoplanets}
\begin{tabular}{lccccc}
\hline
\hline
Name & C$_{\rm Na}$ & C$_{\rm K}$ & $n_{\rm Na}$ & $n_{\rm K}$ & (H$_2$O$-$J)/$H_{\rm eq}$ \\
\hline
Physical units & -- & -- & (cm$^{-3}$) & (cm$^{-3}$) & -- \\
\hline
WASP-6b & $9.8 \pm 8.4$ &$7.7 \pm 3.4$ & 417 & 300 & -- \\
WASP-17b & $2.0 \pm 0.8$ & -- & 208 & -- & $0.67 \pm 0.29$ \\
WASP-31b & -- & $3.7 \pm 1.1$ & -- & 202 & $ 0.86 \pm 0.48$ \\	
WASP-39b & $8.3 \pm 5.9$ & $8.8 \pm 6.9$ & 276 & 200 & -- \\
HAT-P-1b & $5.2 \pm 2.6$ & $8.0 \pm 7.5$ & 376 & 266 & $2.13 \pm 0.61$ \\
HAT-P-12b & $11.0 \pm 10.5$ & $7.9 \pm 6.8$ & 379 & 271 & $0.21 \pm 0.60$ \\
HD 189733b & $10.0 \pm 3.8$ & -- & 656 & -- & $1.86 \pm 0.36$ \\
\hline
\hline
\end{tabular}\\
\end{center}
\end{table*}

\section{Deriving the Cloudiness Index}

We have demonstrated that measuring the spectral slope alone does not allow us to infer the mean molecular mass or the degree of cloudiness in an atmosphere.  Next, we proceed to construct a cloudiness index that does not depend on the spectral slope.

\subsection{Step 0: caveats}

Our cloudiness index is meaningfully defined only if there is a detection of the sodium or potassium line.  We first wish to establish that, when the line is undetected, it is most probably not due to it being dominated by Rayleigh scattering associated with hydrogen molecules.  Rather, the absence of a sodium or potassium line is either due to a vanishingly low abundance of sodium or potassium or a very cloudy atmosphere.  Since there is no way to distinguish between these scenarios without additional information, we will focus on applying our cloudiness index only to objects with reported sodium/potassium line detections.

For the sodium or potassium line center to be unaffected by Rayleigh scattering associated with hydrogen molecules, the relative abundance of sodium or potassium (to H$_2$), by number, has to greatly exceed a threshold value,
\begin{equation}
f_{\rm Na/K} \gg \frac{\sigma_{\rm scat}}{\sigma_0}.
\label{eq:condition}
\end{equation}
The cross section for Rayleigh scattering by molecules is \citep{su05},
\begin{equation}
\sigma_{\rm scat} = \frac{24 \pi^3}{n_{\rm ref}^2 \lambda^4} \left( \frac{n_r^2 - 1}{n_r^2+2} \right)^2 K_\lambda,
\end{equation}
where $n_{\rm ref}$ is a reference number density, $n_r$ is the real part of the index of refraction and $K_\lambda$ is the King factor.  

For molecular hydrogen, we set $K_\lambda=1$ and $n_{\rm ref} = 2.68678 \times 10^{19}$ cm$^{-3}$ and take the refractive index to be \citep{cox}
\begin{equation}
n_r = 1.358 \times 10^{-4} \left[ 1 + 7.52 \times 10^{-3} ~\left(\frac{\lambda}{1 ~\mu\mbox{m}} \right)^{-2} \right] + 1.
\end{equation}
For the sodium/potassium cross sections at line center ($\sigma_0$), we take equation (14) of \cite{heng15}.  The condition in equation (\ref{eq:condition}) becomes $f_{\rm Na/K} \gg (T/1000 \mbox{ K})^{1/2} ~10^{-16}$ for both the sodium and potassium doublets.  Given that the elemental abundances of sodium and potassium in the solar photosphere are $2.0 \times 10^{-6}$ and $1.3 \times 10^{-7}$ \citep{lodders03}, respectively, this condition is expected to possess some generality.  For atmospheres in which this condition holds, the sodium/potassium line center (and the wavelengths near it, in the line wings) is unaffected by Rayleigh scattering.

Other caveats include the assumption of an isothermal atmosphere in hydrostatic equilibrium and our neglect of non-local thermodynamic equilibrium (non-LTE) effects \citep{fortney03}.

\subsection{Step 1: measuring the (isothermal) pressure scale height}

For cloudfree atmospheres, we may directly infer the isothermal pressure scale height using equation (\ref{eq:scaleheight}).  For cloudy atmospheres, this approach is invalid.  Equation (\ref{eq:scaleheight}) is robust in the sense that it is based on measuring differences in the transit radii and wavelengths, but this also makes it blind to whether Rayleigh scattering is associated with molecules or aerosols.

\subsection{Step 2: calculating the cloudfree transit radius difference}

\begin{figure}
\begin{center}
\vspace{-0.2in}
\includegraphics[width=\columnwidth]{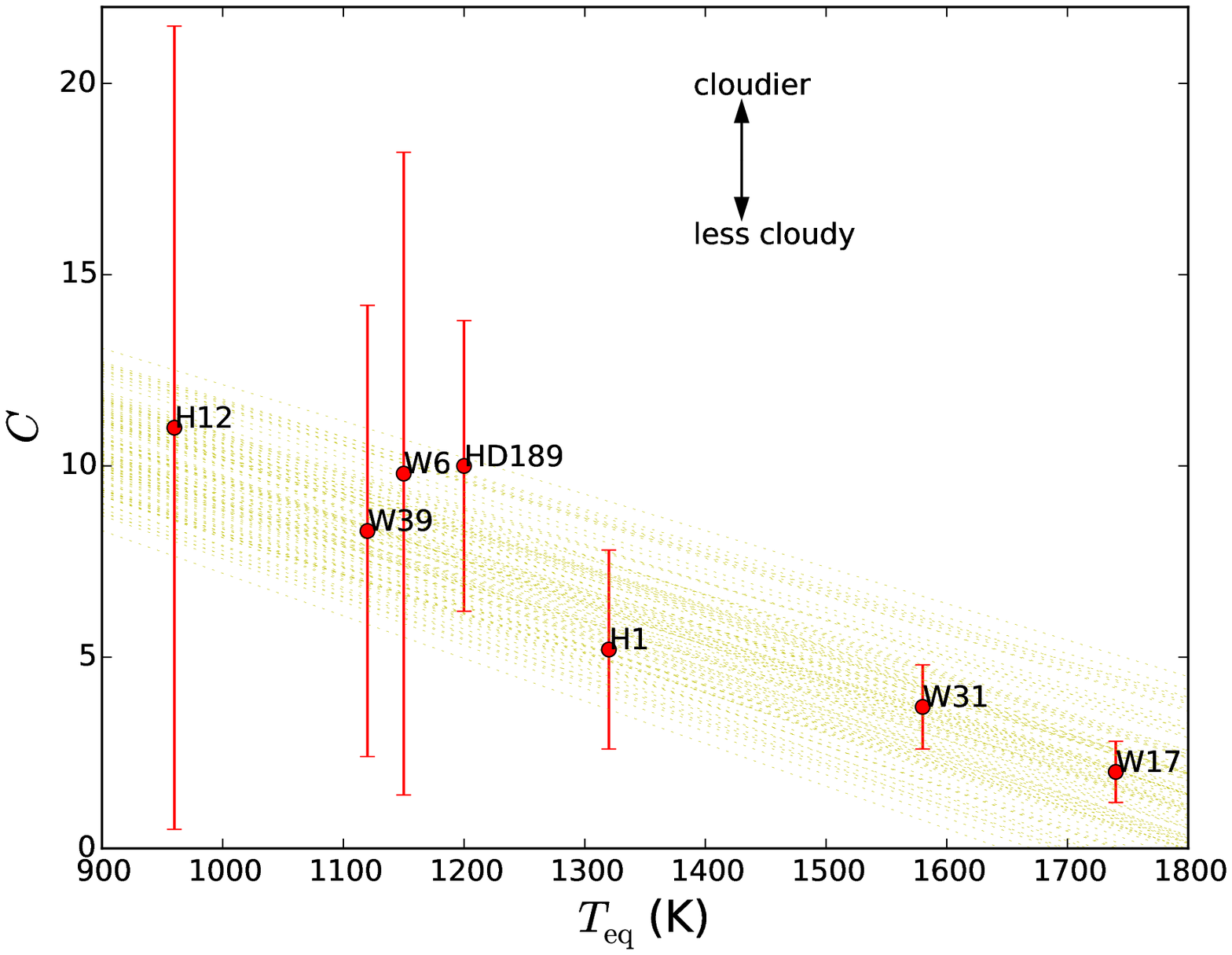}
\includegraphics[width=\columnwidth]{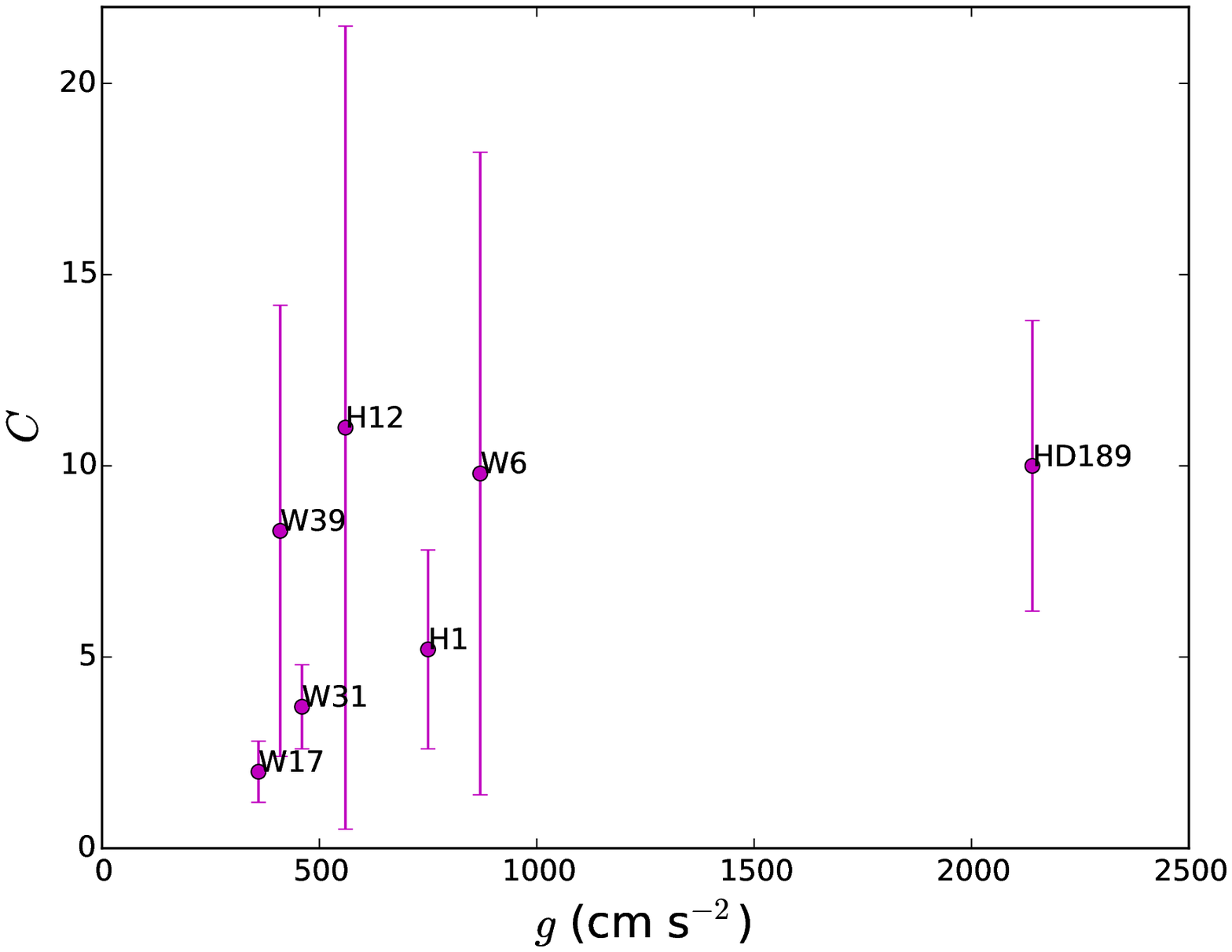}
\includegraphics[width=\columnwidth]{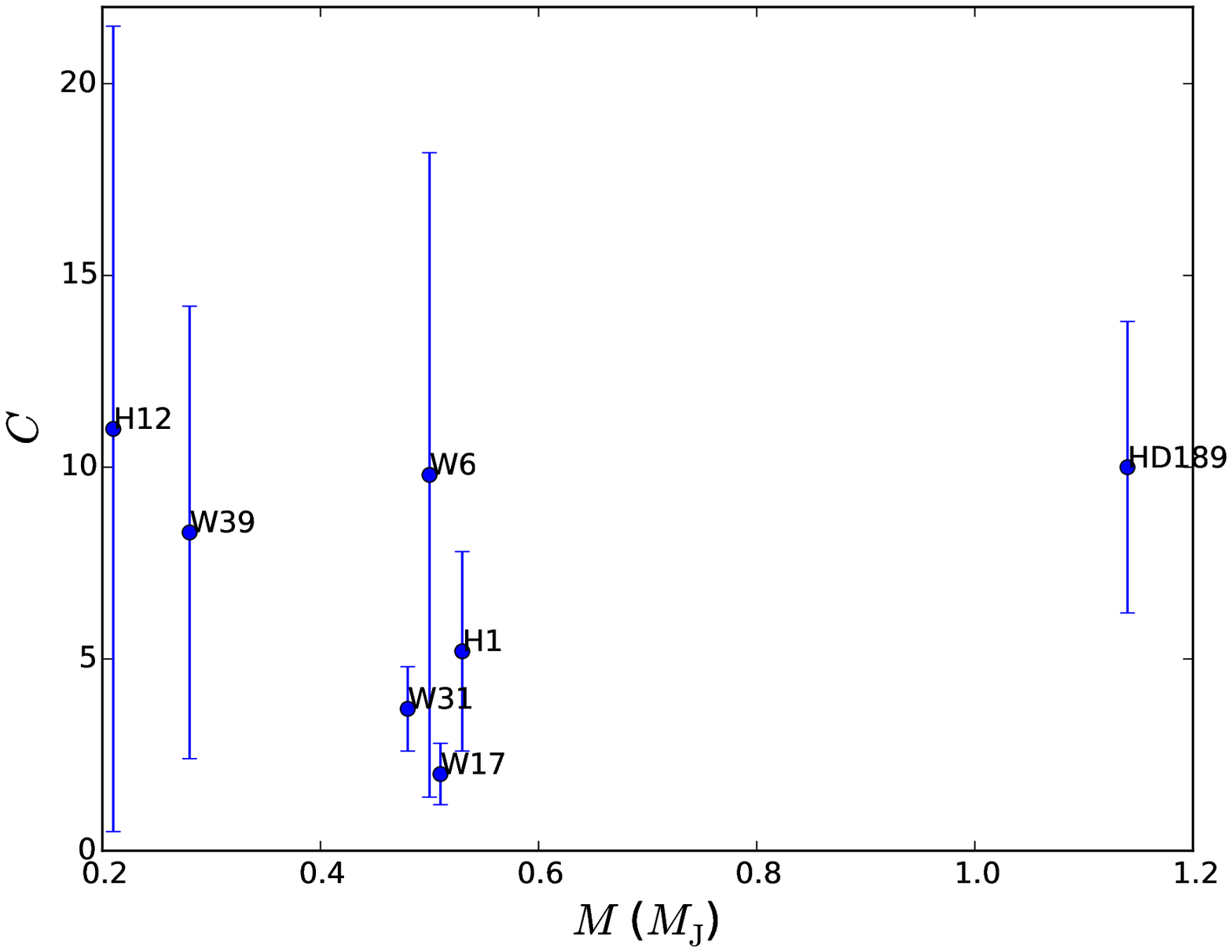}
\end{center}
\caption{Cloudiness index versus equilibrium temperature (top panel), surface gravity (middle panel), exoplanetary mass (bottom panel).  Except for WASP-31b, the values of $C$ are taken from the analysis of the sodium lines.  The labels ``W6", ``W17", ``W31", ``W39", ``H1", ``H12" and ``HD189" refer to WASP-6b, WASP-17b, WASP-31b, WASP-39b, HAT-P-1b, HAT-P-12b and HD 189733b, respectively.  The Spearman rank coefficient for $C$ versus $T_{\rm eq}$, $g$ and $M$ are $-0.86, 0.64$ and $-0.21$, respectively.  The corresponding $p$-values are $0.01, 0.12$ and $0.64$, where the null hypothesis is that the quantities are uncorrelated.  A least-squares linear fit to $C(T_{\rm eq})$ yields coefficients of $(-1.04 \pm 0.10) \times 10^{-2}$ K$^{-1}$ and $(2.00 \pm 0.18) \times 10$ for the linear and constant components, respectively.  The thin dotted lines are 100 Monte Carlo realizations of the linear fit using these coefficients.}
\label{fig:trends}
\end{figure}

In a cloudfree atmosphere, the difference in transit radii, between the line center and wing of the sodium or potassium line, should take on a specific value.  We denote this quantity by $\Delta R$.  Using equations (16) and (17) of \cite{heng15}, we obtain
\begin{equation}
\Delta R = H ~\ln{\left[ \lambda_0 \Phi^{-1} \left( 2 \pi H g \right)^{-1/2} \right]},
\end{equation}
based on the reasoning that the chord optical depths associated with the transits at line center and wing have the same value.  The line-center wavelength is denoted by $\lambda_0$.  The line profile, in the line wings, is well approximated by a Lorentz profile \citep{heng15},
\begin{equation}
\Phi = \frac{A_{21} \lambda^2 \lambda^2_0}{4 \pi c^2 \left( \lambda - \lambda_0 \right)^2},
\end{equation}
where the Einstein A-coefficient is $A_{21}$ and the speed of light is $c$.  For the sodium D$_1$ and D$_2$ lines, we have $A_{21}=6.137 \times 10^7$ s$^{-1}$ and $6.159 \times 10^7$ s$^{-1}$, respectively, corresponding to $\lambda_0 = 0.5897558$ $\mu$m and 0.5891582 $\mu$m \citep{draine11}.  For the potassium D$_1$ and D$_2$ lines, we have $A_{21}=3.824 \times 10^7$ s$^{-1}$ and $3.869 \times 10^7$ s$^{-1}$, respectively, corresponding to $\lambda_0 = 0.770108$ $\mu$m and 0.766701 $\mu$m \citep{draine11}.  For both the sodium and potassium doublets, we estimate that $\Delta R/H \approx 20$ with a gentle dependence on temperature.  It implies that the isothermal pressure scale height need not be known precisely to compute $\Delta R/H$.  In other words, the uncertainty on $\Delta R$ is linearly proportional to the uncertainty on $H$.

A possible concern is that we have not taken pressure broadening of the line into account, which may render our calculations inaccurate.  By using the expression for the chord optical depth \citep{fortney05,heng15} and the ideal gas law, one may obtain an expression for the pressure probed during a transit,
\begin{equation}
P \approx \frac{g}{\kappa} \left( \frac{H}{2\pi R} \right)^{1/2}.
\end{equation}
The preceding expression is essentially the photospheric pressure with a correction term for transit geometry.  Here, $\kappa$ is interpreted as being the mean opacity of the atmosphere.  \cite{freedman14} have shown that the Rosseland and Planck mean opacities are $\sim 0.01$ cm$^2$ g$^{-1}$ and $\sim 1$ cm$^2$ g$^{-1}$, respectively.  If we adopt typical numbers ($g=10^3$ cm s$^{-2}$, $H=1000$ km and $R=R_{\rm J}$, where $R_{\rm J}=7.1492 \times 10^9$ cm is the radius of Jupiter), then we obtain $P \approx 0.05$--$5$ mbar.  We expect the sodium and potassium lines to probe pressures that are similar.  \cite{allard12} have previously shown that pressure broadening of the sodium lines is only important for $P \gtrsim 1$ bar.  Therefore, we conclude that pressure broadening is not a concern.

\subsection{Step 3: measuring the actual difference in transit radii between line center and wing}

By measuring the actual difference in transit radii between the line center and wing ($\Delta R_{\rm obs} \equiv R_0 - R$), we may construct an index for the degree of cloudiness in the atmosphere,
\begin{equation}
C \equiv \frac{\Delta R}{\Delta R_{\rm obs}}.
\end{equation}
Completely cloudfree atmospheres have $C = 1$.  Atmospheres remain nearly cloudfree when $C \sim 1$.  Cloudy atmospheres have $C \gg 1$.

We estimate the uncertainty associated with $C$, denoted by $\delta_C$, using
\begin{equation}
\frac{\delta_C}{C} = \sqrt{\left(\frac{\delta_{H}}{H}\right)^2 + \left(\frac{\delta_{R_0}}{R_0 - R}\right)^2 + \left(\frac{\delta_R}{R_0 - R}\right)^2},
\label{eq:cerror}
\end{equation}
where the uncertainties associated with the transit radii at line center ($R_0$) and wing ($R$) are denoted by $\delta_{R_0}$ and $\delta_R$, respectively.  We do not include the uncertainties on the stellar radius.  

\subsection{Bonus step: measuring the absolute sodium and potassium abundances}

As a bonus, one may directly infer the \textit{absolute} abundance of sodium or potassium associated with the line center \citep{heng15},
\begin{equation}
n_{\rm Na/K} \approx \sqrt{\frac{g}{R_0}} \frac{m_e c}{\pi e^2 f_{\rm osc} \lambda_0},
\end{equation}
where $R_0$ is the transit radius at line center, $m_e$ is the mass of the electron, $e$ is the elementary charge and $f_{\rm osc}$ is the oscillator strength.  For the sodium D$_1$ and D$_2$ lines, we have $f_{\rm osc}=0.32$ and 0.641, respectively \citep{draine11}.  For the potassium D$_1$ and D$_2$ lines, we have $f_{\rm osc}=0.34$ and 0.682, respectively \citep{draine11}.  The preceding expression assumes a cloudfree atmosphere.  In a cloudy atmosphere, it would yield an upper limit for the abundance of sodium or potassium.

\section{Application to Data}

To illustrate the usefulness of the cloudiness index, we estimate its value for a sample of hot Saturns/Jupiters taken from \cite{sing16}.  Table 1 lists the data gleaned from \cite{sing16}\footnote{\texttt{http://www.astro.ex.ac.uk/people/sing \\ /David\_Sing/Spectral\_Library.html}}.  The stellar radii are taken from \cite{johnson08}, \cite{gillon09}, \cite{hartman09}, \cite{anderson11}, \cite{southworth12}, \cite{boya15} and \cite{mac16}.  We augment the Hubble Space Telescope data of \cite{sing16} with ground-based data of the potassium line from \cite{wilson15} for HAT-P-1b.  For the transit radius in the sodium line wing, we choose to extract the data point immediately redward (instead of blueward) of line center to minimize the effects of Rayleigh scattering by molecules.  For the potassium line wing, we extract the data point immediately blueward.

\subsection{Basic analysis}

To begin our analysis requires that we have knowledge of the isothermal pressure scale height.  But as we have already demonstrated, this cannot be reliably inferred in a cloudy atmosphere from measuring the spectral slope.  Instead, we begin by computing the isothermal pressure scale height assuming that the temperature is the equilibrium temperature ($T=T_{\rm eq}$),
\begin{equation}
H_{\rm eq} \equiv \frac{k_{\rm B} T_{\rm eq}}{m g},
\end{equation}
which is then used to compute $\Delta R$ and $C$.  We assume $m=2.4 m_{\rm H}$, where $m_{\rm H}$ is the mass of the hydrogen atom.  In other words, we are assuming that $H=H_{\rm eq}$, following \cite{sing16} and \cite{stevenson16}.  We set the error or uncertainty associated with this assumption to be $\delta_H/H=0.17$ (see \S\ref{subsect:herror}).

Table 2 shows our estimates for $C$, $n_{\rm Na}$ and $n_{\rm K}$.  We subscript $C$ with either ``Na" or ``K", depending on whether it was derived using the sodium or potassium lines---generally, there is consistency between them.  We have taken the average of the values of $A_{21}$ and $f_{\rm osc}$.  Following \cite{sing16}, we omit WASP-12b, WASP-19b and WASP-31b from the analysis for sodium.  We also exclude HD 209458b as there are no data points that precisely align with the sodium line centers.  However, we include WASP-6b and HAT-P-12b.  For the potassium line, we analyze the data of WASP-6b, WASP-31b, WASP-39b and HAT-P-12b, following \cite{sing16}.  We also use the ground-based data of \cite{wilson15} for HAT-P-1b. 

We consistently obtain $n_{\rm Na}, n_{\rm K} \sim 10^2$ cm$^{-3}$.  Only for the nearly cloudfree WASP-17b, WASP-31b and HAT-P-1b are the values of $n_{\rm Na}$ and $n_{\rm K}$ actual estimates for the absolute abundances of sodium and potassium.  For the other objects, they are upper limits as their atmospheres are cloudy.  Note that we cannot estimate the mixing ratios (relative abundances) based on analyzing the sodium or potassium lines alone, because we do not have an independent estimate of the \textit{total} pressure being sensed.  If the atmosphere is hydrogen-dominated (which is our assumption), then this would be the pressure associated with molecular hydrogen.

\subsection{Estimating the error associated with $H=H_{\rm eq}$ assumption}
\label{subsect:herror}

Generally, the temperature being sensed is not the equilibrium temperature, implying that there is an error associated with assuming $H=H_{\rm eq}$.  We can estimate what this error is by calculating the true pressure scale height ($H$) for the nearly cloudfree objects using equation (\ref{eq:scaleheight}) and comparing it to $H_{\rm eq}$.  Specifically, we may perform this analysis for WASP-17b, WASP-31b and HAT-P-1b, since they have $C \sim 1$.  We note that the claim for HAT-P-1b being cloudfree is consistent with the work of \cite{mon15}.   

By specializing to $\alpha=-4$, equation (\ref{eq:scaleheight}) tells us that
\begin{equation}
H = - \frac{1}{4} \frac{\partial R}{\partial \left( \ln{\lambda} \right)}.
\label{eq:scaleheight2}
\end{equation}
We may directly estimate the value of $\partial R /\partial (\ln{\lambda})$ by performing a linear fit to the spectral slopes, in the visible range of wavelengths, measured by \cite{sing16} for WASP-17b, WASP-31b and HAT-P-1b.  We use the data points from the bluest available wavelength up to the data point just blueward of the peak of the sodium line.

For WASP-17b, we have $H_{\rm eq} = 1662$ km.  Our linear fit and use of equation (\ref{eq:scaleheight2}) yields $H= 1896$ km, which translates into an error of $\delta_H/H = 12.3\%$.  For WASP-31b, we obtain $H_{\rm eq}=1181$ km, $H=1390$ km and $\delta_H/H=15.0\%$.  For HAT-P-1b, we obtain $H_{\rm eq}=605$ km, $H=485$ km and $\delta_H/H=24.6\%$.  If we take the average of these three values, we obtain $\delta_H/H \approx 0.17$.  This is the value we assume when using equation (\ref{eq:cerror}).

\subsection{Trends}

Figure \ref{fig:trends} plots $C$ versus the equilibrium temperature, surface gravity and mass ($M$) of the exoplanets.  Curiously, there seems to be a tentative trend of increasing $C$ with decreasing $T_{\rm eq}$, which is a proxy for the incident stellar flux.  The uncertainties associated with $C$ are larger for cooler objects, because of the larger uncertainties on the transit radii.  

By contrast, it has previously been shown that the measured geometric albedos of hot Jupiters exhibit no trend with equilibrium temperature, surface gravity or stellar metallicity, with Kepler-7b being the oddball of having an unusually high geometric albedo \citep{hd13,anger15}.  Physically, if this trend is real, it implies that clouds consisting of sub-micron-sized particles become less prevalent as the atmosphere becomes more irradiated.  Any trend of $C$ with $g$ or $M$ is less apparent.  

\subsection{Comparison with index of Stevenson (2016)}

For 5 out of 7 objects, we may directly compare our values of $C$ with the values of the near-infrared index of \cite{stevenson16}, which is denoted by (H$_2$O$-$J)/$H_{\rm eq}$.  \cite{stevenson16} regards index values of (H$_2$O$-$J)/$H_{\rm eq}>2$ to correspond to cloudfree atmospheres, while index values between 1 and 2 correspond to partially cloudy to nearly cloudfree atmospheres.  Cloudy atmospheres have (H$_2$O$-$J)/$H_{\rm eq}<1$.  One would expect that atmospheres with $C \sim 1$ would also have (H$_2$O$-$J)/$H_{\rm eq} \ge 1$.  Within the uncertainties associated with both indices, HAT-P-1b fulfills the designation of being nearly cloudfree.  Based on both indices, HAT-P-12b is cloudy.  However, the comparison becomes less clear for WASP-17b, WASP-31b and HD 189733b.  It is conceivable that the visible and near-infrared wavelengths are probing different atmospheric layers and one layer being cloudfree does not preclude the other being cloudy.  Future work with a larger sample, where both indices may be calculated, is needed to shed light on these discrepancies.

\section{Summary and Future Work}

We have developed a dimensionless index to quantify the degree of cloudiness of an atmosphere, based on measuring the transit radii at the line center and wing of the sodium and/or potassium lines.  The value of this index has a lower limit of 1, which corresponds to a completely cloudfree atmosphere.  Larger values of the index correspond to cloudier atmospheres.  Physically, our index measures the influence of clouds consisting of sub-micron-sized particles and is complementary to the near-infrared index of \cite{stevenson16}, which probes somewhat larger particles.

We have computed the index for a small sample of 7 hot Saturns/Jupiters taken from \cite{sing16}.  We find a tentative trend of decreasing cloudiness with increasing equilibrium temperature (Figure \ref{fig:trends}).  Future work should measure the sodium and potassium lines at even higher resolutions for a larger sample of exoplanetary atmospheres, in order to confirm or refute this trend.  If the trend holds, it will allow us to screen for a sub-sample of cloudfree objects for further scrutiny by the James Webb Space Telescope, via reconnaissance of a large sample of objects using ground-based measurements.

\acknowledgments
I acknowledge financial support from the Swiss National Science Foundation, the PlanetS NCCR (National Center of Competence in Research) framework and the Swiss-based MERAC Foundation.  I thank the referee for a constructive report.


\label{lastpage}

\end{document}